\documentclass[12pt]{article}

\usepackage{subcaption}

\usepackage{float}

\usepackage{multirow} 

\usepackage{color}
\definecolor{lightblue}{rgb}{0,0.2,0.5}

\usepackage[colorlinks=true, urlcolor=blue,linkcolor=blue, citecolor=lightblue]{hyperref}

\usepackage[round,sort,comma,numbers]{natbib}

\bibpunct{\textcolor{lightblue}{(}}{\textcolor{lightblue}{)}}{,}{a}{}{;}

\usepackage{tikz}

\usetikzlibrary{calc,backgrounds,arrows,matrix}

\usepackage{enumerate}

\usepackage{amssymb,amsmath} 

\usepackage{graphicx} 

\usepackage{graphics} 
\usepackage{color} 

\usepackage{multirow}

\DeclareMathAlphabet{\eufrak}{U}{}{}{} 
\SetMathAlphabet\eufrak{normal}{U}{euf}{m}{n}
\SetMathAlphabet\eufrak{bold}{U}{euf}{b}{n}

\allowdisplaybreaks

 \def\qu{{\mathord{\mathbb Z}}}

 \def\Var{{\mathrm{{\rm Var}}}}

 \def\sZZ{{\rm Z\kern-.45em{}Z}}

 \def\sQQ{{\kern 0.27em \vrule height1.45ex width0.03em depth0em
           \kern-0.30em \rm Q}}
 \def\qu{{\mathchoice
         {\sQQ}
         {\sQQ}
   {\kern 0.225em \vrule height1.05ex width0.025em depth0em \kern-0.25em \rm Q}
   {\kern 0.180em \vrule height0.78ex width0.020em depth0em \kern-0.20em \rm Q}
         }}
 \def\sGG{{\kern 0.27em \vrule height1.45ex width0.03em depth0em
           \kern-0.30em \rm G}}
 \def\gg{{\mathchoice
         {\sGG}
         {\sGG}
   {\kern 0.225em \vrule height1.05ex width0.025em depth0em \kern-0.25em \rm G}
   {\kern 0.180em \vrule height0.78ex width0.020em depth0em \kern-0.20em \rm G}
         }}

 \newtheorem{prop}{Proposition}[section]
 \newtheorem{lemma}[prop]{Lemma}

 \newtheorem{theorem}[prop]{Theorem}
 \newtheorem{remark}[prop]{Remark}

\numberwithin{equation}{section}

 \def\P{{\mathord{\mathbb P}}}

\def\E{\mathop{\hbox{\rm I\kern-0.20em E}}\nolimits}

\makeatletter
\newcommand*\rel@kern[1]{\kern#1\dimexpr\macc@kerna}
\newcommand*\widebar[1]{
  \begingroup
  \def\mathaccent##1##2{
    \rel@kern{0.8}
    \overline{\rel@kern{-0.8}\macc@nucleus\rel@kern{0.2}}
    \rel@kern{-0.2}
  }
  \macc@depth\@ne
  \let\math@bgroup\@empty \let\math@egroup\macc@set@skewchar
  \mathsurround\z@ \frozen@everymath{\mathgroup\macc@group\relax}
  \macc@set@skewchar\relax
  \let\mathaccentV\macc@nested@a
  \macc@nested@a\relax111{#1}
  \endgroup
}
\makeatother

 \newcounter{hyp}
\usepackage{textgreek}

\newenvironment{Proof}{\removelastskip\par\medskip \noindent{\em Proof.} \rm}{\penalty-20\null\hfill$\square$\par\medbreak}

\def\bprf{\begin{Proof}}
\def\nprf{\end{Proof}}
\def\bdes{\begin{description}}
\def\ndes{\end{description}}

\newtheorem{thm}{Theorem}[section]

\def\bdef{\begin{defn}}
\def\ndef{\end{defn}}
\def\bthm{\begin{thm}}
\def\nthm{\end{thm}}
\def\bprop{\begin{prop}}
\def\nprop{\end{prop}}
\def\brmk{\begin{remark}}
\def\nrmk{\end{remark}}
\def\bexa{\begin{exa}}
\def\nexa{\end{exa}}
\def\blem{\begin{lem}}
\def\nlem{\end{lem}}
\def\bcor{\begin{cor}}
\def\ncor{\end{cor}}
\def\bexe{\begin{exe}}
\def\nexe{\end{exe}}

\def\rit{\Bbb{R}}

\def\dit{\Bbb{D}}

\newcommand{\ee}{\mathbb{E}}

\newcommand{\real}{\mathbb{R}}
\newcommand{\pp}{\mathbb{P}}

\makeatletter
\newcommand{\xRightarrow}[2][]{\ext@arrow 0359\Rightarrowfill@{#1}{#2}}
\makeatother

\def\pp{\Bbb{P}}

\def\E{\mathop{\hbox{\rm I\kern-0.20em E}}\nolimits}
\def\Var{\mathop{\hbox{\rm Var}}\nolimits}

\title{\Huge
 A $q$-binomial extension of the CRR asset pricing model 
}

\author{Jean-Christophe Breton\footnote{Univ Rennes, France. 
		Email: jean-christophe.breton@univ-rennes1.fr}
  \and
  Youssef El-Khatib\footnote{United Arab Emirates University, UAE. 
		Email: youssef\_elkhatib@uaeu.ac.ae}
	\and Jun Fan\footnote{University of Nottingham Ningbo, China. Email: Jun.Fan@nottingham.edu.cn} 
	\and Nicolas Privault\footnote{Nanyang Technological University, Singapore. Email: nprivault@ntu.edu.sg}
		}

\allowdisplaybreaks

\begin{document}

\maketitle

\vspace*{-1cm}

\begin{abstract} 
  We propose an extension of the Cox-Ross-Rubinstein (CRR) model based on $q$-binomial (or Kemp) random walks, with application to default with logistic failure rates. This model allows us to consider time-dependent switching probabilities varying according to a trend parameter on a non-self-similar binomial tree.
  In particular, it includes tilt and stretch parameters that control increment sizes. Option pricing formulas are written using $q$-binomial coefficients, and we study the convergence of this model to a Black-Scholes type formula in continuous time. A convergence rate of order $O(N^{-1/2})$ is obtained. 
\end{abstract} 

 \medskip
 

 
\baselineskip0.7cm

\section{Introduction} 
\noindent
 The binomial option pricing model was introduced in
 \cite{sharpe} and established in \cite{crr},
 based on a recombining binary tree 
 allowing for two different market returns $u$, $d$
 at every time step.
 This model leads to tractable option pricing formulas
 for vanilla and exotic options
 using binomial coefficients,
 that converge to the Black-Scholes pricing formula see e.g.
$\S$15-1 of \cite{williams}, 
 and \S~5.7 of \cite{follmerschied}. 
Although pricing and hedging in
the CRR model can be extended to time-dependent parameters, 
 see e.g. \S~11.4 in \cite{privaultbk2},
 the corresponding option pricing formulas
 have exponential instead of polynomial complexity,
 see also \cite{georgiadis}. 

 \medskip

 Extending the flexibility of the CRR model using a wider class
 of parameters while maintaining its original polynomial complexity
 has important consequences for financial modeling. 
{ 
  As the CRR model uses constant switching probabilities over time,
  it does not account possible acceleration effects in economic
  recessions or recoveries. 
  For example, investors may overreact to a bad performance of a stock, 
  and become more sensitive to a decrease in the underlying asset prices, 
  see e.g. \cite{soroka}.
}
 In this paper, we construct a CRR type model based on a $q$-binomial random walk
with time-dependent
switching probabilities, by replacing
standard binomial coefficients with $q$-binomial coefficients.
This provides an alternative to the binomial model
by maintaining its polynomial complexity
and by allowing the underlying security
to move up and down with probabilities increasing or decreasing
according to a trend parameter.
 This feature allows us to model default probabilities
  in a model where default risk can be compounded into
  a failure rate which can increase or decrease over time
  according to a logistic expression. 
  This can for example apply to the modeling of
   accelerating economic recession or expansion phases. 
  In addition, this model allows for the inclusion of a stretch parameter.

\medskip

Convergence of approximations in the continuous-time limit is
another important issue. 
 Convergence rates of the order $O(N^{-1})$ in the number of
 time steps for the price of European call options
 have been obtained in \cite{leisen} for the model of \cite{tian1993},
 see also \cite{diener} for a general asymptotic expansion
 for call options prices in the CRR model. 
 In \cite{tian1999}, a flexible CRR model with an additional ``tilt''
 parameter has been proposed, and extended in \cite{chang}
 as a general class of binomial models with a drift parameter
 that yields smooth convergence at the rate $O(N^{-1})$,  
 with expansions in powers of $1/\sqrt{N}$ for the prices of
 digital and vanilla put and call options.
 In addition, arbitrarily fast convergence in binomial trees has been
 achieved in \cite{joshi2010} and \cite{xiao} respectively for odd and
 even number of time steps,
 and in \cite{leduc} for the model of \cite{chang}.
 
 \medskip
 
 An optimal-drift model further reducing the order
 of convergence of the discretisation error from $O(N^{-1})$ to $o(N^{-1} )$
 has been proposed in \cite{korn-mueller}.
 As noted in \cite{leducbms}, such high-order convergence results
 cannot be reached when the strike price is too
 far away from the spot price. 
 It has also been shown in \cite{leducSAP} that
as the number $N$ of time steps
tends to infinity,
  the order of convergence of tree-based
 methods is in general $O(N^{-1})$ 
 for continuous payoff functions,
 and $O(1/\sqrt{N})$ otherwise. 
 See also \cite{heston-zhou} for related ``local'' convergence rates,
 and \cite{walsh} for
 expressions of the coefficients of $1/\sqrt{N}$ and $1/N$ 
 in the expansion of options prices with general
 payoff functions in a CRR-related model. 

 \medskip 
  
 In \cite{ritchken}, an additional ``stretch'' parameters has been introduced
 for the fine tuning of the pricing of barrier options in the trinomial model.
 In \cite{gcrr} the convergence
 of such a generalized CRR (GCRR) model has been studied,
 and has been shown to be of order $O(1/\sqrt{N})$ in
 the presence of stretch, and of order $O(N^{-1})$ without stretch, see Theorem~2 therein.
 Convergence of order $O(N^{-1})$ is also obtained in \cite{gcrr}
 for a suitable sequence of stretch parameters converging to one,
 see Corollary~1 therein. 

 \medskip 
  
 Our $q$-binomial model also includes an additional parameter $\theta >0$,
 see \eqref{qk} below,
 which plays the role of a stretch parameter.  
 In order to evaluate the continuous-time limit of our model in
option pricing, we will derive convergence results for $q$-binomial
random walks. 
 As noted in \cite{charalambides4},
 the central limit theorem does not hold for the $q$-binomial distribution.
 On the other hand, the convergence of the $q$-binomial distribution
 has been studied for fixed $q \in (0,1)$,
 and convergence to a discrete Heine distribution
 has been shown in \cite{gerhold},
 see also \cite{kyri}.

\medskip 
  
 In our financial setting we consider the case where $q=q_N$ depends
 on the total number of steps $N$,
 and study the convergence of the $q_N$-binomial
 model on $\{0,1,\ldots , N\}$
 when $q_N$ tends to one as $N$ tends to infinity. 
 In this setting, we show in Theorem~\ref{prop:CVtightD2} that
 when $q_N$ takes the form
 $$q_N := 1 + \eta \left(\frac{T}{N} \right)^{3/2} + O(N^{-2})
 $$
 for some $T>0$,
 where the parameter $\eta \in \real$
 captures the ``intensity'' of the trend $q_N$,
 and the market returns $d_N$, $u_N$ satisfy 
\begin{equation}
\nonumber 
 d_N := 
 1 - \sigma \sqrt{\frac{\theta T}{N}} + \zeta \sigma^2 \frac{\theta T}{2N}
 + o( N^{-1} ),
 \quad 
 u_N := 1+ \sigma \sqrt{\frac{\theta T}{N}}
 + \zeta \sigma^2 \frac{\theta T}{2N}
 + o( N^{-1} ), \quad 
\end{equation} 
 this model converges to a continuous-time
 (generalized) Black-Scholes price model
 of the type
 $$
 d\widebar{S}_t = \sigma \widebar{S}_t dB_t
 + \zeta \frac{\sigma^2}{2} \widebar{S}_t dt
 + \frac{\sigma \eta \sqrt{\theta }}{1+\theta } t \widebar{S}_t dt, 
$$
 with time-dependent affine interest rate on $[0,T]$,
 where $(B_t)_{t\in [0,T]}$ is a standard Brownian motion. 
 This result is proved 
 in the sense weak convergence on the Skorohod space
 $\dit([0,T],\rit)$ 
 of c\`adl\`ag functions, equipped with the $J_1$ topology given by the distance 
\begin{equation}
\nonumber 
d_{J_1}(x,y)=\inf_{\lambda\in\Lambda} \big( \|x-(y\circ\lambda)\|_\infty+\|{\rm Id}-\lambda\|_\infty\big), \quad x,y\in \dit([0,T],\rit),
\end{equation}
where ${\rm Id}$ denotes identity, $\|x\|_\infty=\sup_{t\in[0,T]} |x(t)|$,
and $\Lambda$ denotes the set of strictly increasing mappings $\lambda$ of $[0,T]$ onto itself, see \cite[Section 12]{billingsley1999}. 
 In addition, in Theorem~\ref{t51} we show convergence of European
 option prices
  at the rate $O(N^{-1/2})$. 
  For this, we use an expansion for the distribution of sums of
 non identically distributed Bernoulli
 random variables, see Theorem~1.3 in \cite{deheuvels},
 which extends Uspensky's theorem,
 see \S~VII-11 of \cite{uspensky},
 which applied to i.i.d. case. 
 
 \medskip

 We proceed as follows. 
 In Section~\ref{s2} we review the main properties of $q$-binomial
 distributions 
 and we prove a central limit theorem for the associated $q$-binomial random walk. 
 Our $q$-binomial extension of the CRR model is presented
 in Section~\ref{s3}.
 Theorem~\ref{t51} shows the weak convergence of our model
 to a generalized Black-Scholes model using a geometric Brownian motion 
 with affine time-dependent drift,
 with convergence rates of order $O(N^{-1/2})$ or $O(N^{-1})$
 depending on the model parameters. 
 Theorem~\ref{t51} is proved in Sections~\ref{s5}-\ref{s6}  
 by studying the weak convergence
 of the $q$-binomial random walk in continuous time. 
 
\section{The $q$-binomial distribution} 
\label{s2}
This section contains the basic knowledge on $q$-binomial distribution
and random walk that will be needed in the sequel. 
Consider the
 sequence $(X_k)_{k\geq 1}$ of independent Bernoulli random variables with
 variable distribution parameterized in the logistic form 
\begin{equation}
  \label{qk} 
 \P_{\theta , q} (X_k=0) = \frac{1}{1+\theta q^{k-1}}
 \quad
 \mbox{and}
 \quad
 \P_{\theta , q} (X_k=1) = \frac{ \theta q^{k-1}}{1+\theta  q^{k-1}}, 
\end{equation} 
$k \geq 1$, where $\theta , q>0$, see 
\cite{berkson1953}, \cite{cox}. 
This parametrization has been used in \cite{kemp} to construct
the following extension of the binomial distribution with application to 
the statistical study of dice rolling experiments.
 This distribution is also derived in e.g.
 Corollary~3.1 of \cite{charalambides2} and Theorem~9.5 of \cite{charalambides3}. 
 In the next proposition we consider the time-inhomogeneous random walk
 $(Z_n)_{n\geq 0}$ associated to the $q$-binomial distribution. 
 \begin{prop}
  Let $(X_k)_{k\geq 1}$ be a sequence of independent Bernoulli random variables
  with distribution \eqref{qk}.
  The sum
 $$
 Z_n : = X_1+\cdots + X_n, \qquad n \geq 1,  
 $$
 with $Z_0:=0$, has the distribution 
   \begin{equation}
     \label{zn} 
    \P_{\theta , q} ( Z_N - Z_n = k ) 
    =
    \frac{\theta^k q^{(2n+k-1)k/2}
}{(1+\theta q^n)(1+\theta q^{n+1} )\cdots (1+\theta q^{N-1})}
    { N-n \choose k}_q, 
\end{equation} 
 $k = 0,1,\ldots , N-n$, $0\leq n \leq N$, 
 and the probability generating function
$$ 
 \E_{\theta , q} \big[ t^{Z_n} \big] = 
 \frac{(1+\theta t q )\cdots (1+\theta t q^{n-1})}{(1+\theta q)\cdots (1+\theta q^{n-1})},
 \qquad t\in [0,1], 
 $$
  where 
\begin{equation*}
 {n \choose k}_q
 := \frac{(1-q^n)\cdots (1-q^{n-k+1})}{(1-q ) \cdots (1-q^k)},
    \qquad k = 0,1,\ldots , n, 
\end{equation*} 
is the $q$-binomial, or Gaussian binomial, coefficient.
\end{prop}
\begin{Proof}
  For completeness, we provide a proof by induction on $N\geq n$.
  Relation~\eqref{zn} is clearly satisfied for $N=n$ and $N=n+1$.
  Assuming that \eqref{zn} is satisfied at the rank $n\geq N$, we have
\begin{eqnarray*} 
  \lefteqn{
   \! \! \! \! \! \! \!   
    \P_{\theta ,q} (Z_{N+1}-Z_n=k ) 
     = 
  \frac{1}{1+\theta q^N}
  \P_{\theta ,q} (Z_N-Z_n=k ) 
 +
 \frac{\theta q^N}{1+\theta q^N}
 \P_{\theta ,q} (Z_N - Z_n=k-1) 
  }
  \\
  & = &   
\frac{1}{(1+\theta q^N)}
     \frac{ \theta^k q^{(2n+k-1)k/2}
}{(1+\theta q^n)(1+\theta q^{n+1})\cdots (1+\theta q^{N-1})}
    { N-n \choose k}_q
    \\
     & & +
 \frac{\theta q^N}{(1+\theta q^N)}
 \frac{\theta^{k-1} q^{(2n+k-1)(k-2)/2}
}{(1+\theta q^n )(1+\theta q^{n+1} )\cdots (1+\theta q^{N-1})}
          { N-n \choose k-1}_q
 \\
  & = & 
     \frac{    \theta^k q^{(2n+k-1)k/2}
}{(1+\theta q^n )(1+\theta q^{n+1} )\cdots (1+\theta q^N)}
     \left(
          { N-n \choose k}_q
 +
 q^{N-n-(k-1)}
          { N-n \choose k-1}_q
          \right)
 \\
  & = & 
     \frac{    \theta^k q^{(2n+k-1)k/2}
}{(1+\theta q^n )(1+\theta q^{n+1} )\cdots (1+\theta q^N)}
          { N+1-n \choose k}_q, 
\end{eqnarray*} 
where we applied the $q$-Pascal rule
$$
{ N+1-n \choose k}_q
=
{ N-n \choose k}_q
 +
 q^{N-n-(k-1)}
          { N-n \choose k-1}_q, 
          $$
            see Proposition~6.1 in \cite{victor}. 
            Next,
 regarding the probability generating function, we have 
\begin{eqnarray*}
  \E_{\theta , q} \big[ t^{Z_n} \big]
  & = &  
  \sum_{k=0}^n
  t^k
  \P_{\theta , q} ( Z_n = k )
  \\
   & = &  
  \sum_{k=0}^n
       \frac{    ( \theta t)^k q^{(k-1)k/2}
}{(1+\theta )(1+\theta q)\cdots (1+\theta q^{n-1})}
            { n \choose k}_q
              \\
   & = &  
              \left(
              \prod_{l=1}^n \frac{1}{1+\theta q^{l-1}}
              \right)
              \sum_{k=0}^n
        ( \theta t)^k q^{(k-1)k/2}
            { n \choose k}_q
              \\
   & = &  
              \prod_{l=1}^n \frac{1+\theta t q^{l-1}}{1+\theta q^{l-1}}
            , 
\end{eqnarray*} 
 where we used 
 Gauss's binomial formula 
$$
 \sum_{k=0}^n ( \theta t)^k q^{(k-1)k/2}
     { n \choose k}_q
     =
     \prod_{l=1}^n (1+\theta t q^{l-1}),
$$ 
 see Relation~(5.5) in \cite{victor}. 
\end{Proof}
  The $q$-binomial distribution
reduces to the binomial distribution as $q$ tends to $1$,
 since the $q$-binomial coefficients converge to
the standard binomial coefficients when $q$ tends to $1$,
see e.g. Chapter~6 of \cite{victor}.
\subsubsection*{Central limit theorem} 
 As noted in Example~5.5 of \cite{charalambides4},
 the central limit theorem does not hold for the $q$-binomial distribution
 when $q\not=1$, as in this case we have 
 $$
 \lim_{N\to \infty} \Var_{\theta,q} [ Z_N]
 = \lim_{N\to \infty} 
 \sum_{k=1}^N 
 \P_{\theta , q} (X_k=0) \P_{\theta , q} (X_k=1)
 =
 \lim_{N\to \infty} 
 \sum_{k=1}^N 
 \frac{\theta q^{k-1}}{
 (1 + \theta q^{k-1})^2 } 
 < \infty, 
$$
 see, e.g., Theorem 1.1 in \cite{deheuvels}.
 It has been shown in \cite{gerhold} that for fixed $q \in (0,1)$,
 the distribution of $Z_N$ converges to a discrete Heine distribution
 as $N$ tends to infinity, see also \S10.8.1 in \cite{njohnson}. 
 In Theorem~2 of \cite{kyri}, it has been shown that when $\theta $ takes the form 
 $\theta_N = q^{-\vartheta N}$ for some $\vartheta \in (0,1)$,
 the distribution of $Z_N$ can be approximated by a deformed
 Stieltjes-Wigert distribution as $N$ tends to infinity.
 However, this requires 
 $\P_{\theta_N , q} ( Z_n = k )$ to tend to zero for $k=1,\ldots , n-1$,
 which may not be meaningful in
 a financial setting.
 In the sequel, we will study the convergence of the $q$-binomial
 model on $\{1,\ldots , N\}$
 when the parameter $q$ depends on $N$ and tends to one as $N$ tends
 to infinity. 

 \medskip 

 In the next proposition, we show that the $q$-binomial random walk
 converges to a Gaussian distribution under a suitable assumption on 
 $q_N$.
\begin{prop}
\label{p2.2} 
 Assume that $q_N = 1 + O( N^{-3/2})$ as $N$ tends to infinity. 
 Then, letting
 $Z_N : = X_1+\cdots + X_N$,
 $N \geq 1$, the normalized sequence $(Z_N-\E_{\theta , q_N} [Z_N])/\sqrt{N}$ converges
 in distribution to a Gaussian ${\cal N}(0,\theta / ( 1 + \theta )^2 )$ random variable as $N$ tends to infinity. 
\end{prop}
\begin{Proof} 
 For $1 \leq k \leq N$, by the binomial theorem, we have
$$ 
q_N^{k-1} = 1 + (k-1) O( N^{-3/2}) + R^{(k)}_N,
$$
where
\begin{eqnarray*}
  \big| R^{(k)}_N \big| & = & \left|
  \sum_{l=2}^{k-1} {k-1\choose l} ( O( N^{-3/2}) )^l
  \right| 
\\
& = & \sum_{l=0}^{k-3} {k-1\choose l+2} |O( N^{-3(l+2)/2})|
\\
& \leq & \frac{1}{N} \sum_{l=0}^{k-3} {k-3\choose l} |O( N^{-3l/2})|
\\
& = & \frac{1}{N}
 ( 1 +  |O ( N^{-3/2} )| )^{k-3} 
\\
& \leq & \frac{1}{N}
 ( 1 +  |O ( N^{-3/2} )| )^N 
\\
& = & O(N^{-1}). 
\end{eqnarray*} 
 Therefore, we have
$$ 
q_N^{k-1} = 1 + k O ( N^{-3/2} ) + O (N^{-1}),
$$
where the above terms $O ( N^{-3/2} )$ and
$O (N^{-1})$ are uniform in $k\in \{1,\ldots , N\}$,
 hence
\begin{eqnarray} 
  \nonumber
    \P_{\theta , q_N} ( X_k = 1 ) & = & \frac{ \theta q^{k-1}_N}{1+\theta q^{k-1}_N}
  \\
\nonumber
     & = & 
  \theta \frac{
    1+ k O ( N^{-3/2} ) 
    + O (N^{-1}) 
    }{1+\theta + \theta k O ( N^{-3/2} ) + O (N^{-1})
  } 
  \\
\nonumber
    & = &
  \frac{\theta}{1+\theta} 
    \big( 1+ k O ( N^{-3/2} )  
    + O (N^{-1})
    \big) \left( 1-\frac{\theta }{1+\theta } k O ( N^{-3/2} ) 
    + O (N^{-1})
    \right) 
  \\
  \label{fdskjf}
   & = &
  \frac{ \theta}{1+\theta}
  + \frac{k\theta}{(1+\theta)^2}
  O ( N^{-3/2} ) 
 + O (N^{-1}). 
\end{eqnarray} 
 Thus, we have 
\begin{eqnarray*} 
      \E_{\theta , q_N} [ Z_N] & = & \sum_{k=1}^N \P_{\theta , q_N} ( X_k = 1 ) 
      \\
       & = & 
      \frac{\theta }{1+\theta } \sum_{k=1}^N
      \left(
  1 
+ \frac{k}{1+\theta } O ( N^{-3/2} ) 
 + O (N^{-1})
 \right)
\\
       & = & 
\frac{\theta N}{1+\theta } + \frac{\theta O(N^{1/2} )}{2(1+\theta )^2} + O (1). 
\end{eqnarray*}
 The variance of $Z_N$ is then given by 
\begin{align} 
  \nonumber
& \sigma_N^2 : = \Var_{\theta,q_N} [Z_N] 
 = \sum\limits_{k=1}^{N} \P_{\theta , q_N} ( X_k = 0 ) \P_{\theta , q_N} ( X_k = 1 ) 
   \\
 \nonumber
  & =  \sum\limits_{k=1}^{N} 
  \left(  \frac{\theta}{1+\theta} + k O ( N^{-3/2} ) 
+ O (N^{-1})
\right)
\left(  \frac{1}{1+\theta} - k O ( N^{-3/2} ) 
+ O (N^{-1})
\right)
\\
\label{fdskfa}
 & = \frac{\theta N}{(1+\theta)^2} + O ( N^{1/2} )
\end{align} 
 as $N$ tends to infinity, and we conclude by the
 Lindeberg-Feller central limit theorem,
 as in e.g. Theorem 1.1 of \cite{deheuvels}. 
\end{Proof} 
\subsubsection*{$q$-geometric distribution}
 We end this section with some comments on the 
 $q$-geometric distribution
 associated to the $q$-binomial distribution. 
 Let $\tau$ denote the time the first ``$0$'' appears in the sequence
$(X_k)_{k\geq 1}$, i.e.
$$
\tau : = \inf \{ k \geq 1 \ : \ X_k = 0\}.
$$
We have
$$
\P_{\theta , q} ( \tau \geq k )
= \prod_{l=1}^{k-1} \P_{\theta , q} (X_l=1)
= \prod_{l=1}^{k-1} \frac{\theta q^{l-1}}{1+\theta q^{l-1}},
\qquad k \geq 1,
$$
with $\prod_{l=1}^0 = 1$ by convention, and
\begin{eqnarray*} 
\P_{\theta , q} ( \tau = k ) & = & \P_{\theta , q} ( \tau \geq k ) - \P_{\theta , q} ( \tau > k )
\\
 & = & \prod_{l=1}^{k-1} \frac{\theta q^{l-1}}{1+\theta q^{l-1}}
- \prod_{l=1}^k \frac{\theta q^{l-1}}{1+\theta q^{l-1}}
\\
 & = & 
 \frac{1}{1+\theta q^{k-1}}
 \prod_{l=1}^{k-1} \frac{\theta q^{l-1}}{1+\theta q^{l-1}}
, 
\qquad k \geq 1,
\end{eqnarray*} 
 which is the $q$-geometric distribution of the first kind according
to \S~2.10 in \cite{charalambides4},
 and yields the standard geometric distribution with parameter $\theta / ( 1 + \theta )$ when $q=1$.
Taking $\tau$ as default time, this yields the discrete-time
logistic failure rate 
\begin{equation}
  \label{fjkl} 
\P_{\theta , q} ( \tau = k \mid \tau\geq k) = \frac{\P_{\theta , q} ( \tau=k)}{\P_{\theta , q} ( \tau\geq k)}
= \frac{1}{1+\theta q^{k-1}},  \qquad k \geq 1, 
\end{equation} 
see (21) in \cite{coxreg}, (7) in \cite{wathompson}, or (0.8) in \cite{fahrmeir}. 
This rate is constant in the geometric case,
and increasing (resp. decreasing) when $q<1$ (resp. $q>1$). 
\section{A $q$-binomial CRR Model} 
\label{s3}
 Given $d, u$
 such that $0 < d < u$, 
 we consider a risky asset price with initial value $S_0$
 given in discrete time as 
$$ 
   S_n  = S_0 u^{Z_n} d^{n-Z_n}
    = \left\{ 
 \begin{array}{ll} 
uS_{n-1}, & X_n=1, 
\\ 
\\
d S_{n-1}, & X_n=0, 
\end{array} 
\right. 
$$ 
 $n\geq 1$, where $(X_k)_{k\geq 1}$ is as in \eqref{qk}, with the distribution
$$
    \P_{\theta , q}  ( S_n = S_0 u^kd^{n-k} ) 
    =
    \frac{\theta^k q^{(k-1)k/2}
}{(1+\theta )(1+\theta q)\cdots (1+\theta q^{n-1})}
    { n \choose k}_q, \quad k = 0,1,\ldots , n, 
$$
 and 
 \begin{equation}
    \label{n1} 
    \E_{\theta , q} \left[ \frac{S_n }{S_{n-1}} 
  \right]
    = 
    \frac{d +u \theta q^{n-1}}{1+\theta q^{n-1}}, 
    \quad
          \Var_{\theta,q} \left[ \frac{S_n}{S_{n-1}} 
 \right] 
      =
      (u -d )^2
               \Var_{\theta,q} [X_n] 
      =
  \frac{(u -d )^2 \theta q^{n-1}
      }{(1+\theta q^{n-1})^2},
   \end{equation} 
 $n \geq 1$. 
 From \eqref{qk}, we note that the case $q>1$ is modeling an upward market trend,
 while $q<1$ models a downward trend. 
 On the other hand, letting $q$ tend to $1$ recovers the
 standard CRR model with probabilities $1 / (1+\theta )$
 and $\theta /(1+\theta )$.

 \medskip

     Consider now a riskless asset priced $A_0=1$ at time $0$ and  
$$ 
A_n := \prod_{k=1}^n  (1+r_k) 
$$ 
 at time $n\geq 1$, where $r_k$ satisfies 
\begin{equation}
  \label{1rk} 
1+r_k =
    \E_{\theta , q} \left[ \frac{S_k }{S_{k-1}} 
  \right]
    = 
    \frac{d +u \theta q^{k-1}}{1+\theta q^{k-1}}, 
          \qquad
          k \geq 1. 
\end{equation} 
      As a consequence of \eqref{n1},
we find that the discounted price process
$$ 
 \widetilde{S}_n = 
\frac{S_n}{A_n} = S_n \prod_{k=1}^n  
 (1+r_k)^{-1}, \qquad n \geq 0, 
$$ 
 is a martingale with respect to the filtration $({\cal F}_n)_{n \geq 0}$
 generated by $(S_n)_{n \geq 0}$.
 Therefore, $\P_{\theta , q}$ is the unique risk-neutral probability measure
 and therefore the market model is without arbitrage and complete,
 see e.g. Theorems~5.17 and 5.38 in \cite{follmerschied}.
 
 \medskip
 
 In this setting, the arbitrage-free price at time $n = 0,1,\ldots , N$
 of an option with payoff $\phi ( S_N)$ and
 maturity $N$ is given from \eqref{zn} by
\begin{eqnarray*} 
  \lefteqn{
    \left( \prod_{k=n+1}^N (1+r_k)^{-1} \right) 
   \E_{\theta , q} [\phi ( S_N) \mid {\cal F}_n ]
  }
  \\
  & = & 
 \left( \prod_{k=n+1}^N (1+r_k)^{-1} \right) 
 \sum_{k=0}^{N-n}
 \phi ( S_n u^k d^{N-n-k} ) 
 \P_{\theta , q} ( Z_N - Z_n = k )
\\
 & = & 
     \sum_{k=0}^{N-n}
    \theta^k \frac{ q^{(2n+k-1)k/2}
\phi ( S_n u^k d^{N-n-k} ) 
}{(d+\theta u q^n)\cdots (d+\theta uq^{N-1})}
    { N-n \choose k}_q.
\end{eqnarray*} 
     In addition, \eqref{fjkl} shows that the $q$-binomial model has the ability
    to model accelerating economic recession or expansion phases.  

    \medskip
    
    In the sequel we will 
  study the convergence speed of the discrete $q$-binomial
  approximation in the case of European call options
    by
    defining $q_N$, $d_N$, $u_N$, $r_{k,N}$, $0 \leq k \leq N$,
  as 
  \begin{equation}
    \label{1dfjlks} 
  \left\{
  \begin{array}{l}
 \displaystyle
    q_N := 1 + \eta (\Delta t)^{3/2} + O(N^{-2}),
    \medskip
    \\ 
 \displaystyle
  d_N := 
 1 - \sigma \sqrt{\theta \Delta t} + \zeta \sigma^2 \theta \Delta t / 2 
 + o( N^{-3/2} ),
 \medskip
 \\ 
 \displaystyle
 u_N := 1+ \sigma \sqrt{\Delta t / \theta}
 + \zeta \sigma^2 \Delta t / ( 2\theta ) 
 + o( N^{-3/2} ),
 \medskip
 \\
 \displaystyle
 r_{k,N} =
    \frac{d_N -1 +(u_N -1) \theta q_N^{k-1}}{1+\theta q_N^{k-1}}, 
          \qquad
          k \geq 1. 
  \end{array}
  \right.
  \end{equation} 
 As a consequence of Proposition~\ref{p2.2} or Theorem~\ref{prop:CVtightD2} below, 
 it can be shown that the arbitrage-free price
$$
      \left(
      \prod_{k=n+1}^N (1+r_{k,N})^{-1}
      \right)
      \E_{\theta , q} [( S_{N,N} - K)^+ \mid {\cal F}_n ]
   $$
 of a European call option
 with strike price $K$ converges to the continuous-time limit 
\begin{equation}
\label{gfkjldsg}
       \exp \left(
 - \zeta \frac{\sigma^2}{2} T
 - \frac{\sigma \eta T^2 \sqrt{\theta } }{2(1+\theta )}
  \right) 
 \E \big[ \big(\widebar{S}_T - K\big)^+ \big] 
   = S_0 \Phi ( d_+ ) - K
 \exp \left(
  - \zeta \frac{\sigma^2}{2} T
  - \frac{\sigma T^2 \eta \sqrt{\theta } }{2(1+\theta )}
  \right) 
\Phi ( d_- )
\end{equation} 
 as $N$ tends to infinity, where 
$$
d_+ := \frac{1}{\sigma \sqrt{T}} \left( \log \left( \frac{S_0}{K} \right)
+ (1 + \zeta) \frac{\sigma^2 T}{2} 
+ \frac{\sigma \eta T^2 \sqrt{\theta } }{2(1+\theta )}
\right), 
\qquad
d_- = d_+ - \sigma \sqrt{T}, 
$$    
and $\Phi$ is the cumulative distribution function of the
standard normal distribution.

\medskip

\begin{figure}[H]
  \centering
  \hskip-0.4cm
  \begin{subfigure}[b]{0.49\textwidth}
\includegraphics[width=8cm]{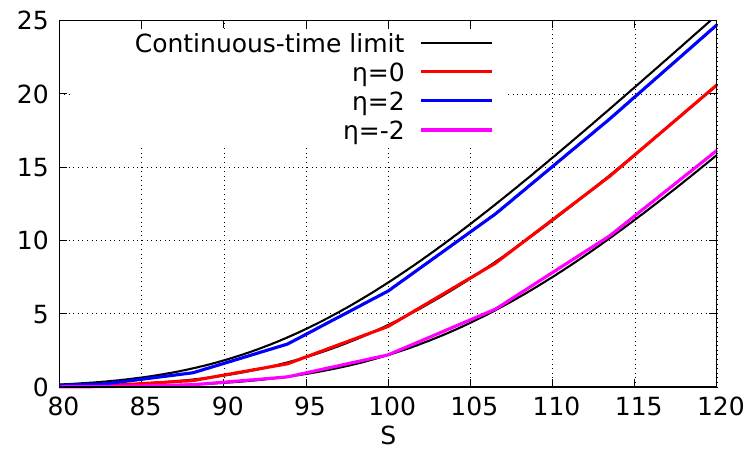} 
\caption{\small Prices with $K=100$, $\sigma = 0.1$.} 
 \end{subfigure}
  \hskip0.4cm
  \begin{subfigure}[b]{0.49\textwidth}
\includegraphics[width=8cm]{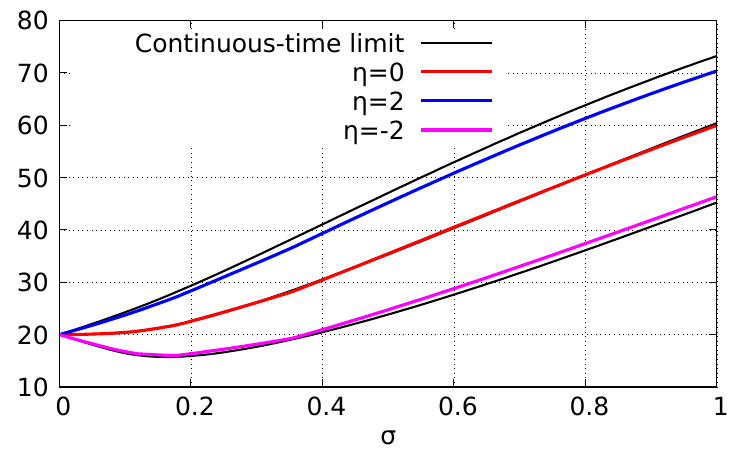} 
\caption{\small Prices with $S=100$, $K=80$.} 
 \end{subfigure}
  \caption{Option price graphs with $\theta = \zeta = 1$.} 
\label{fig1-00} 
\end{figure}

 Figure~\ref{fig1-00} shows European call prices with strike level $K$ 
 as functions of the underlying $S$ and volatility paramter $\sigma$ 
 for $\eta = -2,2$, compared with the CRR and Black-Scholes formulas ($\eta = 0$). 
 We note that option prices may not be monotone functions of
 volatility when $\eta < 0$. 

\vskip-0.4cm

\begin{figure}[H]
  \centering
  \hskip-0.4cm
  \begin{subfigure}[b]{0.49\textwidth}
\includegraphics[width=8cm]{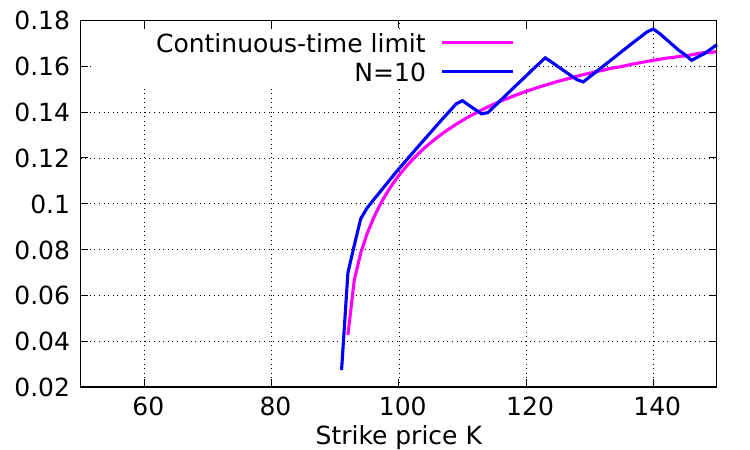} 
\caption{\small Implied volatility for $\eta=-2$.} 
 \end{subfigure}
  \hskip0.4cm
  \begin{subfigure}[b]{0.49\textwidth}
\includegraphics[width=8cm]{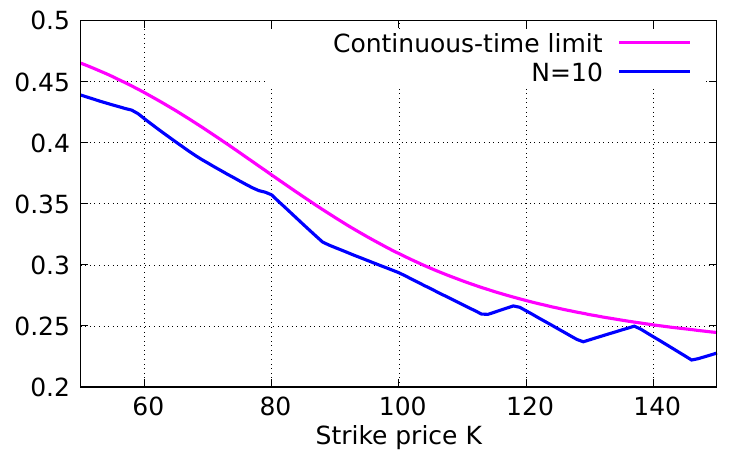} 
\caption{\small Implied volatility for $\eta=2$.} 
 \end{subfigure}
  \caption{Implied volatility graphs with $\theta = \zeta = 1$, $S=100$, $T=1$ and $\sigma = 0.2$.} 
\label{fig1-0} 
\end{figure}

\noindent 
 Figure~\ref{fig1-0} shows the smiles obtained by 
 respectively applying the implied volatilities of the above discrete-time 
 pricing formula and its continuous limit to 
 the CRR and
 (generalized) Black-Scholes formulas \eqref{gfkjldsg} with $\eta = 0$. 
 We note that this composition may not be always defined when $\eta < 0$. 

\vskip-0.4cm

\begin{figure}[H]
  \centering
  \hskip-0.4cm
  \begin{subfigure}[b]{0.49\textwidth}
\includegraphics[width=8cm]{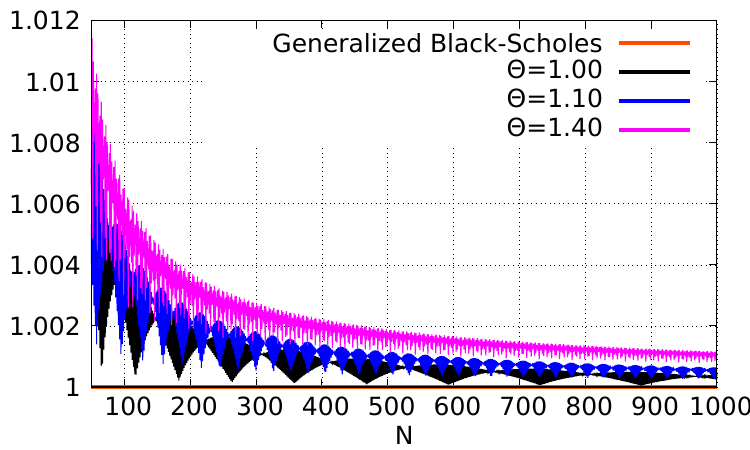} 
\caption{\small Convergence for $\eta=-1$.} 
 \end{subfigure}
  \hskip0.4cm
  \begin{subfigure}[b]{0.49\textwidth}
\includegraphics[width=8cm]{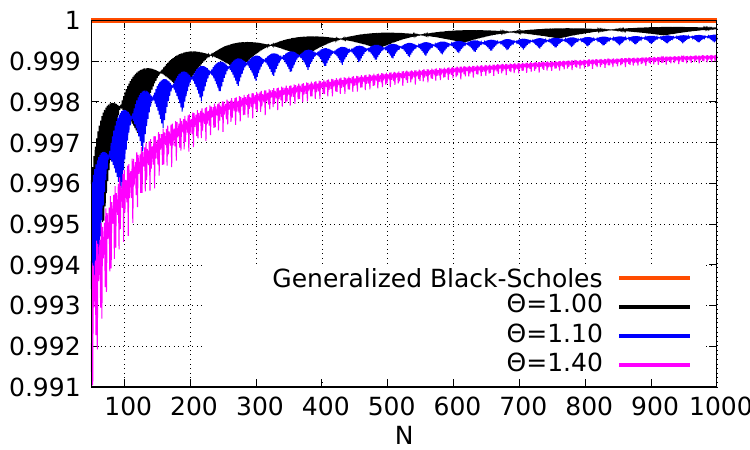} 
\caption{\small Convergence for $\eta=1$.} 
 \end{subfigure}
  \caption{Normalized convergence graphs.} 
\label{fig1} 
\end{figure}

\noindent
The following result, which is proved in Sections~\ref{s5}-\ref{s6},
provides convergence rates for option prices, see also Remark~\ref{r1-1}. 
\begin{theorem} 
  \label{t51}
  Let $r_{k,N}$,
  $d_N$, $u_N$, and $q_N$
  be as in \eqref{1dfjlks}.
  Then, we have
\begin{eqnarray*} 
\lefteqn{ 
        \prod_{k=1}^N ( 1 + r_{k,N})^{-1}
        \E_{\theta , q_N } [ ( S_{N,N} - K)^+ ]
}
\\
& = &
\displaystyle
    S_0 \Phi ( d_+ ) - K
  \exp \left(
  - \zeta  \frac{\sigma^2T}{2}
  - \frac{\sigma \eta T^2 \sqrt{\theta } }{2(1+\theta )}
  \right) 
\Phi ( d_- )
+
O(N^{-1/2}), 
\end{eqnarray*} 
as $N$ tends to infinity.
\end{theorem}
\noindent 
 Figure~\ref{fig1} shows the convergence of option prices normalized to $1$
 with the parameters
 $S_0=100$, $K=95$, $t=0.5$, $\sigma =0.2$, $T=1$ 
 used in \cite{tian1999} and \cite{gcrr}, 
  and $\zeta = 1$. 
 
 Table~\ref{opat0} presents numerical estimates for the graphs of Figure~\ref{fig1}. 

\begin{table}[H]
\centering
\begin{tabular}{|l|l|c|c|c|c|}
\cline{3-5} 
\multicolumn{2}{l|}{}           & \multicolumn{3}{c|}{Number of time steps} &   \multicolumn{1}{l}{}        \\ \hline
\multicolumn{2}{|c|}{Parameters} & $N = 100$ & $N = 1000$ & $N=10000$ & Limit \eqref{gfkjldsg} \\ \hline
\multirow{2}{*}{\begin{tabular}[c] {@{}l@{}} $\eta =1$ \end{tabular}}                                                                  
                                           & $\theta = 1$ & 11.164676 & 11.187615 & 11.189429 & 11.189701 \\ \cline{2-6}
& $\theta = 1.1$ & 1.228880 & 11.253394 & 11.255885 & 11.256045
\\
\hline     \hline     
\multirow{2}{*}{\begin{tabular}[c] {@{}l@{}} $\eta =0$ \end{tabular}}                                                                  
                                           & $\theta = 1$ & 8.949356 & 8.947683 & 8.947027 & 8.947041 \\ \cline{2-6}
& $\theta = 1.1$ & 8.960038 & 8.947035 & 8.947104 & 8.947041 \\
\hline     \hline     
\multirow{2}{*}{\begin{tabular}[c] {@{}l@{}} $\eta =-1$ \end{tabular}}                                                                  
                                           & $\theta = 1$ & 7.008068 & 6.993345 & 6.991934 & 6.991621 \\ \cline{2-6}
                                           & $\theta = 1.1$ & 7.027543 & 6.995757 & 6.994490 & 6.993759 \\ \hline     
\end{tabular}
\caption{Convergence table.}
\label{opat0}
\end{table}
\vspace{-0.4cm}

\section{Weak convergence in continuous time} 
\label{s5}
In the sequel we let $T>0$ denote a terminal time horizon, 
and we use the discretization $\Delta t := T/N$
of the time axis.
Given $\sigma , \theta > 0$ and $\zeta \in \real$, we assume that 
  \begin{equation}
    \label{jfkl} 
 d_N := 
 1 - \sigma \sqrt{\theta \Delta t} + \zeta \frac{\sigma^2}{2} \theta \Delta t
 + o( N^{-1} ) 
 \quad 
 \mbox{and} 
 \quad 
 u_N := 1+ \sigma \sqrt{\frac{\Delta t}{\theta} }
 + \zeta \frac{\sigma^2}{2\theta } \Delta t
 + o( N^{-1} ). 
\end{equation} 
 In the next proposition we show the convergence of
 the $q$-binomial model to 
 a (generalized) continuous-time Black-Scholes model with affine time-dependent 
 risk-free interest rate 
 $$
 r_t :=
 \zeta \frac{\sigma^2}{2}
 + \frac{\sigma \eta \sqrt{\theta}}{1+\theta } t , \qquad t>0, 
$$ 
 as the number $N$ of time steps tends to infinity.
 In the sequel, we let $S_{0,N}:=S_0$ and 
 $$
 S_{k,N} := S_0 u_N^{Z_k} d_N^{k-Z_k}, \qquad k = 1,\ldots , N, 
 $$
 and let
 $\lfloor x \rfloor$ denote the integer part of
 $x \geq 0$, i.e.
 the greatest integer less than or equal to $x$.  

 \medskip

 In the next proposition, we prove that the
stepwise interpolation $S_{\lfloor Nt/T \rfloor ,N}$, $t\in [0,T]$, 
between the random variables $S_{k,N}$, $k=0, \ldots , N$, converges 
to the solution $\big(\widebar{S}_t\big)_{t\in [0,T]}$  of the stochastic differential equation 
$$
 d\widebar{S}_t = \sigma \widebar{S}_t dB_t
 + \zeta \frac{\sigma^2}{2} \widebar{S}_t dt
 + \frac{\sigma \eta \sqrt{\theta }}{1+\theta } t \widebar{S}_t dt, 
$$
 where $(B_t)_{t\in [0,T]}$ is a standard Brownian motion. 
 The choice of the power $3/2$ in \eqref{fjkdl-2}
 is necessary in order for the next result to hold. 
\begin{theorem}
\label{prop:CVtightD2}
 Let $d_N$, $u_N$ be as in \eqref{jfkl}, 
 and assume that $q_N$ depends on $N$ as 
 \begin{equation}
   \label{fjkdl-2} 
 q_N := 1 + \eta (\Delta t)^{3/2} + O(N^{-2}), 
\end{equation} 
 where $\eta \in \real$ and $\Delta t : = T/N$. 
 Then we have the weak convergence 
\begin{equation}
\nonumber 
\left(S_{\lfloor Nt/T\rfloor ,N}\right)_{t\in [0,T]}
\xRightarrow{\dit([0,T],\rit)}
\left(
S_0\exp\left( \frac{\sigma\eta \sqrt{\theta }}{2(1+\theta)}t^2-(1-\zeta)\frac {\sigma^2}{2}t +\sigma B_t\right) \right)_{t\in[0,T]}
\end{equation}
 in $\dit([0,T],\rit)$ as $N$ tends to infinity. 
\end{theorem}
\begin{Proof}
  We start by proving one-dimensional convergence, followed by
  finite-dimensional convergence, and we conclude by showing
  the tightness of the sequence $\left(S_{\lfloor Nt/T \rfloor ,N}\right)_{t\in [0,T]}$.
  We note that weak convergence can be obtained as a consequence of
  one-dimensional convergence for time-homogeneous Markov processes, see e.g.
  Theorems~2.5-2.6 of \cite{Ethier-Thomas-2005},
  however, we prefer to present a self-contained proof here, as our random
  walk is not time-homogeneous. 
  First, we show that for any $t\in [0,T]$ the one-dimensional convergence
  \begin{equation}
    \label{fsdkjfs}
  S_{\lfloor Nt/T \rfloor ,N}
  \xRightarrow
  \displaystyle S_0 \exp\left(\frac{\sigma\eta \sqrt{\theta}}{2(1+\theta)}t^2-(1-\zeta)\frac {\sigma^2}{2}t+\sigma  B_t\right)
  \end{equation}
  holds in
  distribution
   as $N$ tends to infinity.  
 We note that  
\begin{eqnarray}
\nonumber
\lefteqn{\frac{u_N}{d_N}
=
\left( 1+ \sigma \sqrt{\frac{\Delta t}{\theta} }
 + \zeta \frac{\sigma^2}{2\theta } \Delta t
 + o( N^{-1} )\right)\left(1 - \sigma \sqrt{\theta \Delta t} + \zeta \frac{\sigma^2}{2} \theta \Delta t
 + o( N^{-1} ) \right)^{-1}
}
\\
\nonumber
 &=&
\left( 1+ \sigma \sqrt{\frac{\Delta t}{\theta} }
 + \zeta \frac{\sigma^2}{2\theta } \Delta t
 + o( N^{-1} )\right)
 \left(1 + \sigma \sqrt{\theta \Delta t} +\Big(1-\frac \zeta 2\Big) \sigma^2 \theta \Delta t + o( N^{-1} ) \right)
\\
 \label{eq:loguNdN-0}
&=&
1 + \sigma \frac{\theta+1}{\sqrt{\theta}}\sqrt{\Delta t} 
+\sigma^2(1+\theta)\left(1+\zeta \frac{1-\theta}{2\theta}\right)\Delta t 
+ o( N^{-1} ), 
\end{eqnarray}
 hence 
 \begin{equation}
    \label{eq:loguNdN}
 \log \frac{u_N}{d_N}
 =\sigma \frac{\theta+1}{\sqrt{\theta}}\sqrt{\Delta t} 
+\sigma^2\frac{\theta^2-1}{2\theta}(1-\zeta)\Delta t
+ o( N^{-1}),
\end{equation} 
and, letting 
$\displaystyle Z_{\lfloor Nt/T \rfloor} := \sum_{k=1}^{\lfloor Nt/T \rfloor} X_k$, 
by \eqref{fdskjf}-\eqref{fdskfa} we have 
\begin{eqnarray*} 
      \E_{\theta , q_N} [ Z_{\lfloor Nt/T \rfloor}] & = & \sum_{k=1}^{\lfloor Nt/T \rfloor} \pp_{\theta , q_N} ( X_k = 1 ) 
      \\
       & = & 
      \frac{\theta }{1+\theta } \sum_{k=1}^{\lfloor Nt/T \rfloor}
 \left( 1 
 + \frac{\eta k}{1+\theta }  (\Delta t )^{3/2} 
 + O (N^{-1})
 \right)
\\
       & = & 
\frac{\theta}{1+\theta }{\lfloor Nt/T \rfloor} + \frac{\theta \eta}{2(1+\theta )^2} \big(\lfloor Nt/T \rfloor(\lfloor Nt/T \rfloor-1)\big)(\Delta t )^{3/2} + O (1) 
\\
  & = & 
\frac{\theta}{1+\theta }{\lfloor Nt/T \rfloor} +\frac{\theta \eta t^2N^{1/2}}{2(1+\theta )^2\sqrt{T}} + O(1),
\end{eqnarray*}
 and 
\begin{align} 
  \nonumber
&\Var_{\theta,q_N} \big[ Z_{\lfloor Nt/T \rfloor} \big]  
 = \sum_{k=1}^{\lfloor Nt/T \rfloor} \pp_{\theta , q_N} ( X_k = 0 ) \pp_{\theta , q_N} ( X_k = 1 ) 
   \\
 \nonumber
  & =  \sum_{k=1}^{\lfloor Nt/T \rfloor}
  \left(  \frac{\theta}{1+\theta} + \frac{\theta \eta k }{(1+\theta)^2}
  (\Delta t )^{3/2} 
+ O (N^{-1})
\right)
\left(  \frac{1}{1+\theta} - \frac{\theta \eta k}{(1+\theta )^2}
  (\Delta t )^{3/2} 
+ O (N^{-1})
\right)
\\
\nonumber 
 & = \frac{\theta {\lfloor Nt/T \rfloor}}{(1+\theta)^2} + O ( N^{1/2} ) 
 =\frac{\theta t}{T(1+\theta)^2} N + O ( N^{1/2} ) .
\end{align} 
As a consequence, we have   
\begin{align}
\nonumber
& \log S_{\lfloor Nt/T \rfloor ,N}
\\
\nonumber
 & =
\log(S_0)+ \lfloor Nt/T \rfloor \log d_N+Z_{\lfloor Nt/T \rfloor}\log(u_N/ d_N) 
\\
\nonumber
&=\log(S_0)+ \lfloor Nt/T \rfloor \log d_N+\E_{\theta, q_N}\left[Z_{\lfloor Nt/T \rfloor}\right]\log(u_N/ d_N)
\\
\nonumber
&\hspace{5cm} +\big(Z_{\lfloor Nt/T \rfloor}-\E_{\theta, q_N}\left[Z_{\lfloor Nt/T \rfloor}\right]\big)\log(u_N/ d_N)
\\
\nonumber
&=\log(S_0)
+\lfloor Nt/T \rfloor \left(-\sigma \sqrt{\theta \Delta t} + (\zeta-1) \frac{\sigma^2}{2} \theta \Delta t + o( N^{-1} ) \right)
\\
\nonumber
&+\left(\frac{\theta}{1+\theta }{\lfloor Nt/T \rfloor} +\frac{\theta \eta t^2N^{1/2}}{2(1+\theta )^2\sqrt{T}} + O (1)\right)
\left(
\sigma \frac{\theta+1}{\sqrt{\theta}}\sqrt{\Delta t} 
+\sigma^2\frac{\theta^2-1}{2\theta}(1-\zeta)\Delta t
+ o( N^{-1})\right)
\\
\nonumber
&+\big(Z_{\lfloor Nt/T \rfloor}-\E_{\theta, q_N}\left[Z_{\lfloor Nt/T \rfloor}\right]\big)
\left(
\sigma \frac{\theta+1}{\sqrt{\theta}}\sqrt{\Delta t} 
+\sigma^2\frac{\theta^2-1}{2\theta}(1-\zeta)\Delta t
+ o( N^{-1})\right)
\\
\label{eq:esperanceLN}
&=\log(S_0)
+(\zeta-1) \frac{\sigma^2}{2} \theta t + \frac{\sigma^2}2(\theta-1)(1-\zeta)t
+\frac{\eta t^2 \sqrt{\theta} }{2(1+\theta)} \sigma +o(1) 
\\
\nonumber
&+\frac{Z_{\lfloor Nt/T \rfloor}-\E_{\theta, q_N}\left[Z_{\lfloor Nt/T \rfloor}\right]}{\sqrt{\Var_{\theta,q_N} \big[ Z_{\lfloor Nt/T \rfloor} \big] }}
\left(\frac{\theta t}{T(1+\theta)^2} N + O ( N^{1/2} ) \right)^{1/2}
\\
\nonumber
 & \qquad \qquad \times \left(
\sigma \frac{\theta+1}{\sqrt{\theta}}\sqrt{\Delta t} 
+\sigma^2\frac{\theta^2-1}{2\theta} ( 1 - \zeta ) \Delta t
+ o( N^{-1})
\right). 
\end{align}
 By Slutsky's lemma and the Lindeberg-Feller central limit theorem, 
 see \cite[Th. 27.2]{billingsley} or \cite[Th. 7.2.1]{chungbk},
 for the triangular array 
 $(X_k)_{1\leq k\leq \lfloor Nt/T \rfloor}$
 of Bernoulli random variables, 
 since
 $$\lim_{N\to+\infty}\sum_{k=1}^{\lfloor Nt/T \rfloor}
 \pp_{\theta , q_N} ( X_k = 0 ) \pp_{\theta , q_N} ( X_k = 1 )
 =+\infty,
 $$ 
 we have 
$$
\frac{Z_{\lfloor Nt/T \rfloor}-\E_{\theta, q_N}\left[Z_{\lfloor Nt/T \rfloor}\right]}{\sqrt{\Var_{\theta,q_N} \big[ Z_{\lfloor Nt/T \rfloor} \big] }}
\xRightarrow[N\to+\infty]{} {\cal N}(0,1),
$$
hence
$\log S_{\lfloor Nt/T \rfloor ,N}$
 converges in distribution to 
 $$
 {\cal N}\left(
\log(S_0)
+(\zeta-1) \frac{\sigma^2}{2} t
+\frac{\eta \sqrt{\theta} }{2(1+\theta)} \sigma t^2, \sigma^2 t\right),
$$
which implies \eqref{fsdkjfs} by the continuous mapping theorem. 
 More generally, this argument yields the convergence of 
$$
\log \frac{S_{\lfloor Nt_i/T \rfloor ,N}}{S_{\lfloor Nt_{i-1}/T \rfloor ,N}} 
=
\left(\lfloor Nt_i/T \rfloor-\lfloor Nt_{i-1}/T \rfloor\right) \log d_N+\left(Z_{\lfloor Nt_i/T \rfloor}-Z_{\lfloor Nt_{i-1}/T \rfloor}\right)\log \frac{u_N}{d_N}
$$
 in distribution to 
$$
{\cal N}\left((\zeta-1) \frac{\sigma^2}{2} (t_i-t_{i-1})
+\frac{\eta \sqrt{\theta} }{2(1+\theta)} \sigma (t_i^2-t_{i-1}^2), \sigma^2 ( t_i-t_{i-1} ) \right),
$$ 
for any $i=1,\ldots , p$,
 where the term $(t_i^2-t_{i-1}^2)$ above is obtained from the limit of 
\begin{eqnarray*} 
\lefteqn{\E_{\theta , q_N} \big[ Z_{\lfloor Nt_i/T \rfloor}-Z_{\lfloor Nt_i/T \rfloor}\big]
=\frac{\theta }{1+\theta }
\left(\left\lfloor Nt_i/T \right\rfloor-\left\lfloor Nt_{i-1}/T \right\rfloor\right)}
\\&&
+ \frac{\eta }{2(1+\theta)}
\left(\left\lfloor Nt_i/T \right\rfloor \left(\left\lfloor Nt_i/T \right\rfloor-1\right)-
\left(
\left\lfloor Nt_{i-1}/T \right\rfloor
\left(\left\lfloor Nt_{i-1}/T \right\rfloor-1\right)
\right)\right) + o (N^{1/2}).
\end{eqnarray*}
Next, by the independence of the $X_k$'s, the independence of the increments of $(B_t)_{t\in [0,T]}$, and the continuous mapping lemma applied twice, first with the exponential function and second with $g_p(x_1,\dots, x_p)=\big(\prod_{i=1}^kx_i\big)_{1\leq k\leq p}$,
we note that 
for any $p\geq 1$ and $0 = t_0 \leq t_1\leq \cdots\leq t_p\leq T$, the $p$-dimensional random vector
\begin{equation}
\nonumber 
\left(S_{\lfloor Nt_1/T \rfloor ,N}, S_{\lfloor Nt_2/T \rfloor ,N}, \dots,  S_{\lfloor Nt_p/T\rfloor ,N}\right) 
\end{equation}
 converges in distribution to 
\begin{equation}
\nonumber 
\left(S_0 \exp\left(\frac{ \sigma\eta \sqrt{\theta}}{2(1+\theta)}t_i^2+(\zeta-1)\frac {\sigma^2}{2}t_i
+\sigma B_{t_i}\right)\right)_{i=1, \dots, p}
\end{equation}
as $N$ tends to infinity.
Finally, by continuity of the exponential and the limit 
\begin{eqnarray*}
  \lefteqn{
   \! \! \! \! \! \! \! \! \! \! \! \! \! \! \! \! \! \! \! \! \! \! \! \! \! \! \!
    \lim_{N\to \infty}
  \E_{\theta,q_N}[\log S_{\lfloor Nt/T \rfloor ,N}]
   = 
  \lim_{N\to \infty}
  \log(S_0)+\lfloor Nt/T \rfloor \log d_N+\ee[Z_{\lfloor Nt/T \rfloor} ] \log \frac{u_N}{d_N} 
  }
  \\ 
  & = & 
\log S_0+(\zeta-1) \frac{\sigma^2}{2} \theta t 
+\left(\frac{\sigma^2}2(\theta-1)(1-\zeta)t
+\frac{\eta t^2 \sqrt{\theta} }{2(1+\theta)} \sigma\right),
\end{eqnarray*}
 obtained from \eqref{eq:esperanceLN}, 
 we can conclude the proof by showing
 the tightness of the sequence
 $$( L_N(t))_{t\in [0,T]} :=\left(\log S_{\lfloor Nt/T \rfloor ,N}-\E_{\theta,q_N}[\log S_{\lfloor Nt/T \rfloor ,N}]\right)_{t\in [0,T]}.
 $$ 
For this, we note that 
from Lemma~\ref{fjkldsf-l} and the bound $\big(\log (u_N/d_N)\big)^4 \leq C/N^2$
obtained from \eqref{eq:loguNdN} we have 
\begin{eqnarray*}
\E_{\theta,q_N}^x\big[
\big( L_N(t) - L_N(0)\big)^4\big]
&=& \big(\log (u_N/d_N)\big)^4 \E_{\theta,q_N}^x\left[\left(\sum_{k=1}^{\lfloor Nt/T \rfloor} (X_k-\E_{\theta,q_N}[X_k])\right)^4\right]
\\
&\leq&\frac C{N^2}{\lfloor Nt/T \rfloor}^2\leq C \frac{t^2}{T^2},
\end{eqnarray*}
 and we conclude with the tightness criterion of \cite{aldous1978}
 as in \cite[Prop. 34.9]{bassbk2}, see also page~176 of \cite{billingsley}.
\end{Proof}
 The following lemma has been used in the proof of Theorem~\ref{prop:CVtightD2}. 
\begin{lemma}
  \label{fjkldsf-l}
  Let $(Y_n)_{n\geq 1}$ be a sequence of independent and centered random variables with
  uniformly bounded fourth moments : $\ee[Y_n^4]\leq K$, for all $n\geq 1$. 
Then, there exists a finite constant $C$ such that
\begin{equation}
\label{eq:boudsum4}
\ee\Bigg[\left(\sum_{k=1}^nY_k\right)^4\Bigg]
\leq CKn^2.
\end{equation}
\end{lemma}
\begin{Proof}
We have
\begin{eqnarray*}
\left(\sum_{i=1}^nY_i\right)^4 
&=&\sum_{i=1}^n Y_i^4
+{4\choose 2}{2\choose 2} \sum_{\begin{subarray}{c} i,j=1\\i\not=j\end{subarray}}^n Y_i^2Y_j^2
+{4\choose 3}{1\choose 1} \sum_{\begin{subarray}{c} i,j=1\\i\not=j\end{subarray}}^n Y_i^3Y_j
\\
&&+{4\choose 2}{2\choose 1}{1\choose 1}\sum_{\begin{subarray}{c} i,j,k=1\\i\not=j\not=k\end{subarray}}^n Y_i^2Y_jY_k
+{4\choose 1}{3\choose 1}{2\choose 1}{1\choose 1}\sum_{\begin{subarray}{c} i,j,k,l=1\\i\not=j\not=k\not=l\end{subarray}}^n Y_iY_jY_kY_l. 
\end{eqnarray*}
Since the random variables $Y_i$'s are independent and centered, we have  
\begin{eqnarray*}
\ee\Bigg[\left(\sum_{i=1}^nY_i\right)^4\Bigg] 
&=&\sum_{i=1}^n \ee[Y_i^4]+{4\choose 2} \sum_{\begin{subarray}{c} i,j=1\\i\not=j\end{subarray}}^n\ee[Y_i^2]\ee[Y_j^2]\\
&\leq & n K+6n(n-1)K, 
\end{eqnarray*}
since 
$\big(\ee[Y_i^2]\big)^2\leq \ee[Y_i^4]\leq K$. 
The bound \eqref{eq:boudsum4} follows. 
\end{Proof}
\section{Proof of Theorem~\ref{t51}} 
\label{s6} 
\begin{Proof}
 Letting $\theta_N := \theta u_N / d_N$ 
 and denoting by $m$ the smallest integer such that
      $S_0 (u_N)^m (d_N)^{N-m} 
            > K$, 
                        we have 
      \begin{eqnarray} 
        \nonumber
        \lefteqn{
     \! \!    \! \!  \! \! \! \! \! \! \! \! \! \! \! \! \! \! \!
          \!  \! \! \! \!  \! \! \! \! \! \! \!
          \prod_{k=1}^N ( 1 + r_{k,N})^{-1}
        \E_{\theta , q_N} [ ( S_{N,N} - K)^+ ]
 =  S_0
     \prod_{k=1}^N ( 1 + r_{k,N})^{-1}
     \sum_{k=m}^N 
     \frac{   \theta^k u_N^k d_N^{N-k}
       q_N^{(k-1)k/2}
}{(1+\theta )\cdots (1+ \theta q_N^{N-1})}
         { N \choose k}_{q_N}
        }
        \\
\nonumber
                 & & - K  
         \prod_{k=1}^N ( 1 + r_{k,N})^{-1}
         \sum_{k=m}^N 
    \frac{ \theta^k q_N^{(k-1)k/2}
}{(1+\theta )\cdots (1+ \theta q_N^{N-1} )}
         { N \choose k}_{q_N}
         \\
\nonumber
        & = & 
     S_0
     \sum_{k=m}^N 
     \frac{ \theta_N^k 
       q_N^{(k-1)k/2}
}{(1+\theta_N )\cdots (1 +\theta_N q_N^{N-1} )}
         { N \choose k}_{q_N}
         \\
        \nonumber
         & & - K  
         \prod_{k=1}^N ( 1 + r_{k,N})^{-1}
         \sum_{k=m}^N 
    \frac{ \theta^k  q_N^{(k-1)k/2}
}{(1 +\theta )\cdots (1 + \theta q_N^{N-1} )}
         { N \choose k}_{q_N}
         \\
         \label{dfjsk1}
         & = & 
         S_0 \P_{\theta_N,q_N} ( Z_N \geq m )
         - K  \P_{\theta,q_N} ( Z_N \geq m ) \prod_{k=1}^N ( 1 + r_{k,N})^{-1},  
                 \end{eqnarray} 
      where $\P_{\theta_N,q_N} (Z_N =k )$ satisfies \eqref{zn}. 
      By Theorem~1.3 in \cite{deheuvels}, we have
             \begin{align} 
         \nonumber       
 & 
                 \P_{\theta_N,q_N} ( Z_N \geq m ) 
    =  
               1 -     \P_{\theta_N,q_N} ( Z_N \leq m-1 )
                              \\
      \label{de}
 & =  
      \Phi ( -z_m)
      - \frac{1-z_m^2}{6 \sigma_N} \varphi ( z_m)
      \left( 1 - \frac{2}{\sigma_N^2} 
      \sum_{k=1}^N ( \P_{\theta_N , q_N} (X_k=1) )^2 \P_{\theta_N , q_N} (X_k=0)
 \right)
      + O(\sigma_N^{-2}), 
\end{align} 
             where
             $\varphi$ denotes the standard normal probability density function, 
             $\sigma_N^2 = \Var_{\theta_N , q_N} [ Z_N]$, 
             and 
             \begin{align*} 
 z_m & := \frac{ ( m-1 ) -\E_{\theta_N , q_N} [Z_N] + 1/2}{\sqrt{\Var_{\theta_N,q_N} [Z_N]}}
 \\
  & =  
 \frac{  m - \sum_{k=1}^N
   \P_{\theta_N , q_N} (X_k=1) -1/2}{\sqrt{\Var_{\theta_N,q_N} [Z_N]}}
 \\
  & =  
 \frac{      \log ( K / S_0 ) - \sum_{k=1}^N \big( \log d_N  +
   \P_{\theta_N , q_N} (X_k=1)
   \log ( u_N / d_N ) \big)
   }{\log ( u_N / d_N )\sqrt{\Var_{\theta_N,q_N} [Z_N]}}
 +
 \frac{ \varepsilon_N -1/2}{\sqrt{\Var_{\theta_N,q_N} [Z_N]}}, 
\end{align*} 
             with
               \begin{equation} 
\label{fkdsf-00}
0\leq \varepsilon_N := m - \frac{\log ( K / (S_0 (d_N)^N ))}{\log (u_N/d_N)} \leq 1,
\end{equation} 
 since $m$ is the smallest integer such that
      $m > \log (K/(S_0 (d_N)^N )) / \log (u_N/d_N) $.
 Next, by \eqref{eq:loguNdN-0} we have
  \begin{equation} 
\label{fkdsf}
  \theta_N 
= 
\theta + ( 1 + \theta ) \sigma \sqrt{\theta \Delta t} + O (N^{-1}), 
\end{equation} 
 hence 
\begin{equation} 
  \label{fdskjf-2}
  \P_{\theta_N , q_N} (X_k=1)
   = \frac{\theta_N q_N^{k-1}}{1+\theta_N q_N^{k-1}}
   =
   \frac{ \theta}{1+\theta}
 + \frac{\eta \theta k }{(1+\theta )^2} (\Delta t)^{3/2}  
 + \frac{\sigma \sqrt{\theta \Delta t}}{1+\theta} + o (N^{-1/2})
\end{equation} 
 see also \eqref{fdskjf} and \eqref{1dfjlks}. 
 Now, by \eqref{jfkl} and \eqref{eq:loguNdN}, we have 
$$ 
 \log d_N =
 - \sigma \sqrt{\theta \Delta t}
 - (1-\zeta )
 \theta \sigma^2 \frac{\Delta t}{2}
 + o(N^{-1}), 
 \quad
 \log u_N =
 \sigma \sqrt{\frac{\Delta t}{\theta }}
 - (1-\zeta ) \sigma^2 \frac{\Delta t}{2\theta }
 + o(N^{-1}), 
$$ 
 and
$$ 
 \log \frac{u_N}{d_N} =
 \left( \sqrt{\theta} + \frac{1}{\sqrt{\theta}} \right) \sigma \sqrt{\Delta t}
 - ( 1 - \zeta ) \sigma^2 \frac{1- \theta^2}{2\theta}
 \Delta t
 + o(N^{-1}), 
$$ 
 hence 
\begin{align} 
  \nonumber
  &     \sum_{k=1}^N \left( \log d_N  + \P_{\theta_N , q_N} (X_k=1) \log \frac{u_N}{d_N} \right)
=
 \sum_{k=1}^N \Bigg( - \sigma \sqrt{\theta \Delta t}
 - (1-\zeta ) \frac{\theta}{2 } \sigma^2 \Delta t
   \\
\nonumber
   & 
 \quad +
 \frac{1}{1+\theta}
 \left(
 \theta + \frac{\eta \theta k }{1+\theta } (\Delta t)^{3/2}
 + \sigma \sqrt{\theta \Delta t} + o (N^{-1/2})
\right)
 \\
\nonumber
   & 
 \quad
 \quad
 \times \left(
  \left( \frac{1+\theta }{\sqrt{\theta}} \right)
  \sigma \sqrt{\Delta t} 
  +
  ( 1- \zeta ) \frac{\sigma^2 \Delta t}{2} \left( \frac{\theta^2 - 1}{\theta} \right)
  +
  o(N^{-1})
\right)
\Bigg)
\\
\nonumber
   & = 
 \sum_{k=1}^N \Bigg( 
 ( - 1 + \zeta ) \frac{\sigma^2 \Delta t}{2} 
 + \frac{\sigma \eta k \sqrt{\theta}}{1+\theta }(\Delta t)^2
 +
  o(N^{-1})
 \\
\nonumber
   & 
 \quad
 +
 \frac{ \sigma \sqrt{\theta \Delta t} + o (N^{-1/2})}{1+\theta}
 \left(
  \left( \frac{1+\theta }{\sqrt{\theta}} \right)
  \sigma \sqrt{\Delta t} 
  +
  ( 1- \zeta ) \frac{\sigma^2 \Delta t}{2} \left( \frac{\theta^2 - 1}{\theta} \right)
  +
  o(N^{-1})
\right)
\Bigg)
\\
\label{jfklds12}
& = ( 1 + \zeta) \frac{\sigma^2 T}{2}
+ \frac{ \sigma \eta T^2 \sqrt{\theta }}{2(1+ \theta ) }
+ O(N^{-1/2}). 
\end{align}
 Similarly, by \eqref{fdskfa}, \eqref{fjkdl-2} and \eqref{eq:loguNdN}, we have 
\begin{eqnarray}
\nonumber
 \Var_{\theta_N,q_N} [Z_N] 
 \left( \log \frac{u_N}{d_N} \right)^2
&=&
\left( \frac{\theta N}{(1+\theta)^2}
 + o ( N^{1/2} )
 \right)
 \left( \left( \sqrt{\theta}  +\frac{1}{\sqrt{\theta}}\right)^2 \sigma^2 \Delta t
+ O (N^{-3/2}) \right)
\\
\nonumber 
&=& \sigma^2 T + O (N^{-1/2}), 
\end{eqnarray}
 hence 
\begin{eqnarray*} 
    z_m & = & \frac{\log ( K/S_0)
- (1 + \zeta ) \sigma^2 T/2 - \sigma \eta T^2 \sqrt{\theta }/(2(1+ \theta ))
        }{\sigma \sqrt{T}}
      +
 \frac{ \varepsilon_N -1/2}{\sqrt{N \theta_N /(1+\theta_N )^2}}
 + O(N^{-1/2})
 \\
 & = & -d_+  +  \frac{ \varepsilon_N -1/2}{\sqrt{N \theta_N /(1+\theta_N )^2}}
 + O(N^{-1/2}).
\end{eqnarray*} 
    By \eqref{de}, we find 
    \begin{equation} 
      \label{fjlf43} 
        \P_{\theta_N,q_N} ( Z_N \geq m )
= 
      \Phi \left(
      d_+
      - \frac{ \varepsilon_N -1/2}{\sqrt{N \theta_N / ( 1 + \theta_N )^2}}
 + O(N^{-1/2})
 \right) 
 + \frac{1 - \theta}{1+\theta} O(N^{-1/2}) 
      + O(N^{-1}), 
     \end{equation} 
    since by \eqref{fdskjf}-\eqref{fdskfa}, \eqref{fjkdl-2}
    and \eqref{fkdsf} we have 
      $$
      1 - \frac{2}{\sigma_N^2}
      \sum_{k=1}^N ( \P_{\theta_N , q_N} (X_k=1) )^2 \P_{\theta_N , q_N} (X_k=0) 
      = \frac{1 - \theta}{1+\theta}
      + O(N^{-1/2}), 
      $$
      as $z_m$ and $\varphi (z_m)$ are bounded in $N\geq 1$. 
      Repeating the above analysis by replacing $\theta_N$ with $\theta$,
      i.e. replacing \eqref{fdskjf-2} with \eqref{fdskjf}, 
      changing $1+\zeta$ into $1-\zeta$ in \eqref{jfklds12} 
      and $\theta_N$ into $\theta$ in \eqref{fjlf43}, shows that 
      \begin{equation}
        \label{fjdl34} 
        \P_{\theta,q_N} ( Z_N \geq m )
= 
      \Phi \left(
      d_-
      - \frac{ \varepsilon_N -1/2}{\sqrt{N \theta / ( 1 + \theta )^2 }}
+ O (N^{-1/2} ) 
\right) 
+ \frac{1 - \theta}{1+\theta} O(N^{-1/2}) 
      + O(N^{-1}). 
\end{equation}
 Finally, we note that by \eqref{fdskjf} we have 
 \begin{align*}
 & 
    r_{k,N}
 = d_N +( u_N - d_N ) \pp_{\theta,q_N} (X_k=1)-1
    \\
&=
 - \sigma \sqrt{\theta \Delta t} + \zeta \frac{\sigma^2}{2} \theta \Delta t
 + o( N^{-3/2} ) 
 \\
 & 
 +
 \left(
 \sigma \sqrt{\frac{\Delta t}{\theta} }
 + \sigma \sqrt{\theta \Delta t}
 + \zeta \frac{\sigma^2}{2\theta } \Delta t
 - \zeta \frac{\sigma^2}{2} \theta \Delta t
 + o( N^{-3/2} )
 \right)
 \left( 
 \frac{ \theta}{1+\theta}
 + \frac{\eta \theta k }{(1+\theta )^2} (\Delta t)^{3/2}  
 + O (N^{-1})
 \right)
\\
&=\frac{\sigma\eta}{{1+\theta}} k \sqrt{\theta\Delta t} ( \Delta t)^{3/2}+\zeta\frac{\sigma^2\Delta t}2 + O(N^{-3/2}) 
\end{align*}
and 
\begin{eqnarray}
\nonumber
  \sum_{k=n+1}^N r_{k,N}
 & = & \sum_{k=n+1}^N 
\left(\sigma\eta\sqrt{\theta\Delta t} \frac k{1+\theta} (\Delta t)^{3/2}+\zeta\frac{\sigma^2\Delta t}2+O(N^{-3/2}) \right)
\\
\nonumber
&=&\sigma\eta\frac{\sqrt{\theta\Delta t}}{1+\theta} (\Delta t)^{3/2} \sum_{k=n+1}^N k 
+\zeta\frac{\sigma^2\Delta t}2 ( N - n )
+O(N^{-1/2}) 
\\
\nonumber
&=&\sigma\eta\frac{ T^2 \sqrt{\theta }}{2(1+\theta)} 
+\zeta\frac{\sigma^2T}{2} + O(N^{-1/2}), 
\end{eqnarray}
as $N$ tends to infinity.
We conclude from the above identity combined
with \eqref{dfjsk1}, \eqref{fjlf43} and \eqref{fjdl34}. 
\end{Proof}
\begin{remark}
\label{r1-1}
    When $\theta = 1$ 
    we have 
    $$ 
\P_{\theta_N,q_N} ( Z_N \geq m )
=
      \Phi \left(
      d_+
      - \frac{ \varepsilon_N -1/2}{\sqrt{N \theta_N / ( 1 + \theta_N )^2}}
\right) 
+ O(N^{-1})
$$
and
$$
\P_{\theta ,q_N} ( Z_N \geq m )
 = 
      \Phi \left(
      d_-
      - \frac{ \varepsilon_N -1/2}{\sqrt{N \theta / ( 1 + \theta )^2}}
\right) 
      + O(N^{-1}). 
      $$
      If in addition $\zeta=1$, then 
$$
S_0 \P_{\theta_N,q_N} ( Z_N \geq m )
         - K    \exp \left(
  - \frac{ \sigma^2T}{2}
  - \frac{\sigma \eta T^2 }{4}
  \right) 
\P_{\theta,q_N} ( Z_N \geq m ) 
$$
converges
to $\displaystyle  S_0 \Phi ( d_+ ) - K
  \exp \left(
  - \frac{\sigma^2T}{2}
  - \frac{\sigma \eta T^2 \sqrt{\theta } }{2(1+\theta )}
  \right) 
\Phi ( d_- )
$ with rate $O(N^{-1})$ since the contribution of
$( \varepsilon_N -1/2)/\sqrt{N \theta_N / (1+\theta_N )}$, resp.
 $( \varepsilon_N -1/2)/\sqrt{N \theta / (1+\theta )}$, 
is of order $O(N^{-1})$ by \eqref{fkdsf-00}
 because the first derivative of
    $$
    x \mapsto
 S_0 \Phi ( d_+ + x ) - K
  \exp \left(
  - \frac{ \sigma^2T}{2}
  - \frac{\sigma \eta T^2 }{4}
  \right) 
\Phi ( d_- + x )
$$
vanishes at $x=0$. 
\end{remark}

\subsubsection*{Conclusion}
This paper proposes an extension of the CRR model where the standard binomial coefficients are replaced by $q$-binomial  coefficients,
{allowing for the modeling of a compounded default risk at increasing or decreasing rates}. The proposed model uses a random walk with time-dependent probabilities, allowing for greater flexibility without losing the original polynomial complexity of the CRR model. In particular, the underlying asset price can move up and down with probabilities increasing or decreasing according to a trend parameter.
 We have derived a pricing formula for vanilla options generalizing the original CRR option pricing formula,
 and proved the convergence in distribution of our model to a (generalized)
 Black-Scholes type model with time-dependent interest rate. 
 This convergence utilizes a geometric Brownian motion with time-dependent affine drift,
 and it holds with a $O(N^{-1/2})$ rate for European call options.
 
\subsubsection*{Acknowledgment}
We thank the anonymous reviewers for suggestions that helped us improving this paper. 

\footnotesize

\def\cprime{$'$} \def\polhk#1{\setbox0=\hbox{#1}{\ooalign{\hidewidth
  \lower1.5ex\hbox{`}\hidewidth\crcr\unhbox0}}}
  \def\polhk#1{\setbox0=\hbox{#1}{\ooalign{\hidewidth
  \lower1.5ex\hbox{`}\hidewidth\crcr\unhbox0}}} \def\cprime{$'$}


\begin{thebibliography}{43}
\providecommand{\natexlab}[1]{#1}
\providecommand{\url}[1]{\texttt{#1}}
\expandafter\ifx\csname urlstyle\endcsname\relax
  \providecommand{\doi}[1]{doi: #1}\else
  \providecommand{\doi}{doi: \begingroup \urlstyle{rm}\Url}\fi

\bibitem[Aldous(1978)]{aldous1978}
D.~Aldous.
\newblock Stopping times and tightness.
\newblock \emph{Ann. Probability}, 6\penalty0 (2):\penalty0 335--340, 1978.

\bibitem[Bass(2011)]{bassbk2}
R.F. Bass.
\newblock \emph{Stochastic Processes}.
\newblock Cambridge Series in Statistical and Probabilistic Mathematics.
  Cambridge University Press, 2011.

\bibitem[Berkson(1953)]{berkson1953}
J.~Berkson.
\newblock A statistically precise and relatively simple method of estimating
  the bio-assay with quantal response, based on the logistic function.
\newblock \emph{Journal of the American Statistical Association}, 48\penalty0
  (263):\penalty0 565--599, 1953.

\bibitem[Billingsley(1995)]{billingsley}
P.~Billingsley.
\newblock \emph{{P}robability and {M}easure}, volume 245 of \emph{Wiley Series
  in Probability and Statistics}.
\newblock Wiley, third edition, 1995.

\bibitem[Billingsley(1999)]{billingsley1999}
P.~Billingsley.
\newblock \emph{Convergence of {P}robability {M}easures}.
\newblock Wiley series in Probability and Statistics. Wiley-Interscience, 2nd
  edition, 1999.

\bibitem[Chang and Palmer(2007)]{chang}
L.-B. Chang and K.J. Palmer.
\newblock Smooth convergence in the binomial model.
\newblock \emph{Finance Stoch.}, 11:\penalty0 91--105, 2007.

\bibitem[Charalambides(2010)]{charalambides2}
Ch.A. Charalambides.
\newblock Discrete $q$-distributions on {B}ernoulli trials with a geometrically
  varying success probability.
\newblock \emph{J. Statist. Plann. Inference}, 140:\penalty0 2355--2383, 2010.

\bibitem[Charalambides(2016)]{charalambides4}
Ch.A. Charalambides.
\newblock \emph{Discrete {$q$}-distributions}.
\newblock John Wiley \& Sons, Inc., Hoboken, NJ, 2016.

\bibitem[Charalambides(2019)]{charalambides3}
Ch.A. Charalambides.
\newblock A review of the basic discrete {$q$}-distributions.
\newblock In \emph{Lattice path combinatorics and applications}, volume~58 of
  \emph{Dev. Math.}, pages 166--193. Springer, Cham, 2019.

\bibitem[Chung(1995)]{chungbk}
K.L. Chung.
\newblock \emph{A course in probability theory}.
\newblock Wiley Series in Probability and Statistics. Wiley, third edition,
  1995.

\bibitem[Chung and Shih(2007)]{gcrr}
S.-L. Chung and P.-T. Shih.
\newblock Generalized {C}ox-{R}oss-{R}ubinstein binomial models.
\newblock \emph{Management Science}, 53\penalty0 (3):\penalty0 508--520, 2007.

\bibitem[Cox(1958)]{cox}
D.R. Cox.
\newblock The regression analysis of binary sequences.
\newblock \emph{Journal of the Royal Statistical Society: Series B (Statistical
  Methodology)}, 20:\penalty0 215--232, 1958.

\bibitem[Cox(1972)]{coxreg}
D.R. Cox.
\newblock Regression models and life-tables.
\newblock \emph{Journal of the Royal Statistical Society: Series B (Statistical
  Methodology)}, 34\penalty0 (2):\penalty0 187--220, 1972.

\bibitem[Cox et~al.(1979)Cox, Ross, and Rubinstein]{crr}
J.C. Cox, S.A. Ross, and M.~Rubinstein.
\newblock Option pricing: A simplified approach.
\newblock \emph{Journal of Financial Economics}, 7:\penalty0 87--106, 1979.

\bibitem[Deheuvels et~al.(1989)Deheuvels, Puri, and Ralescu]{deheuvels}
P.~Deheuvels, M.L. Puri, and S.S. Ralescu.
\newblock Asymptotic expansions for sums of nonidentically distributed
  {B}ernoulli random variables.
\newblock \emph{J. Multivariate Anal.}, 28:\penalty0 282--303, 1989.

\bibitem[Diener and Diener(2004)]{diener}
F.~Diener and M.~Diener.
\newblock Asymptotics of the price oscillations of a {E}uropean call option in
  a tree model.
\newblock \emph{Math. Finance}, 14:\penalty0 271--293, 2004.

\bibitem[Ethier and Kurtz(2005)]{Ethier-Thomas-2005}
S.N. Ethier and T.G. Kurtz.
\newblock \emph{Markov processes, characterization and convergence}.
\newblock Wiley Series in Probability and Statistics. John Wiley \& Sons Inc.,
  New York, 2005.

\bibitem[Fahrmeir(1997)]{fahrmeir}
L.~Fahrmeir.
\newblock Discrete failure time models.
\newblock Discussion Paper 91, Collaborative Research Center 386, LMU
  M{\"u}nchen, 1997.

\bibitem[F{\"o}llmer and Schied(2004)]{follmerschied}
H.~F{\"o}llmer and A.~Schied.
\newblock \emph{Stochastic finance}, volume~27 of \emph{de Gruyter Studies in
  Mathematics}.
\newblock Walter de Gruyter \& Co., Berlin, 2004.

\bibitem[Georgiadis(2011)]{georgiadis}
E.~Georgiadis.
\newblock Binomial options pricing has no closed-form solution.
\newblock \emph{Algorithmic Finance}, 1\penalty0 (1):\penalty0 13--16, 2011.

\bibitem[Gerhold and Zeiner(2010)]{gerhold}
S.~Gerhold and M.~Zeiner.
\newblock Convergence properties of {K}emp's $q$-binomial distribution.
\newblock \emph{Sankhya: The Indian Journal of Statistics}, 72-A:\penalty0
  331--343, 2010.

\bibitem[Heston and Zhou(2000)]{heston-zhou}
S.~Heston and G.~Zhou.
\newblock On the rate of convergence of discrete-time contingent claims.
\newblock \emph{Math. Finance}, 10\penalty0 (1):\penalty0 53--75, 2000.

\bibitem[Johnson et~al.(2005)Johnson, Kemp, and Kotz]{njohnson}
N.L. Johnson, A.W. Kemp, and S.~Kotz.
\newblock \emph{Univariate discrete distributions}.
\newblock Wiley Series in Probability and Statistics. Wiley-Interscience,
  Hoboken, NJ, third edition, 2005.

\bibitem[Joshi(2010)]{joshi2010}
M.S. Joshi.
\newblock Achieving higher order convergence for the prices of {E}uropean
  options in binomial trees.
\newblock \emph{Mathematical Finance}, 20\penalty0 (1):\penalty0 89--103, 2010.

\bibitem[Kac and Cheung(2002)]{victor}
V.~Kac and P.~Cheung.
\newblock \emph{Quantum Calculus}.
\newblock Universitext. Springer, New York, second edition, 2002.

\bibitem[Kemp and Kemp(1991)]{kemp}
A.W. Kemp and C.D. Kemp.
\newblock Weldon's dice data revisited.
\newblock \emph{Amer. Statist.}, 45\penalty0 (3):\penalty0 216--222, 1991.

\bibitem[Korn and M{\"u}ller(2013)]{korn-mueller}
R.~Korn and S.~M{\"u}ller.
\newblock The optimal-drift model: an accelerated binomial scheme.
\newblock \emph{Finance and Stochastics}, 17:\penalty0 135--160, 2013.

\bibitem[Kyriakoussis and Vamvakari(2013)]{kyri}
A.G. Kyriakoussis and M.~Vamvakari.
\newblock A $q$-analogue of the {S}tirling formula and a continuous limiting
  behaviour of the $q$-binomial distribution - numerical calculations.
\newblock \emph{Methodol. Comput. Appl. Probab.}, 15:\penalty0 187--213, 2013.

\bibitem[Leduc(2013)]{leduc}
G.~Leduc.
\newblock A {E}uropean option general first-order error formula.
\newblock \emph{ANZIAM J.}, 54\penalty0 (4):\penalty0 248--272, 2013.

\bibitem[Leduc(2016{\natexlab{a}})]{leducSAP}
G.~Leduc.
\newblock Option convergence rate with geometric random walks approximations.
\newblock \emph{Stochastic Analysis and Applications}, 34\penalty0
  (5):\penalty0 767--791, 2016{\natexlab{a}}.

\bibitem[Leduc(2016{\natexlab{b}})]{leducbms}
G.~Leduc.
\newblock Can high-order convergence of {E}uropean option prices be achieved
  with common {CRR}-type binomial trees?
\newblock \emph{Bulletin of the Malaysian Mathematical Sciences Society},
  39\penalty0 (4):\penalty0 1329--1342, 2016{\natexlab{b}}.

\bibitem[Leisen and Reimer(1996)]{leisen}
D.P.J. Leisen and M.~Reimer.
\newblock Binomial models for option valuation - examining and improving
  convergence.
\newblock \emph{Appl. Math. Finance}, 3\penalty0 (4):\penalty0 319--346, 1996.

\bibitem[Privault(2009)]{privaultbk2}
N.~Privault.
\newblock \emph{Stochastic analysis in discrete and continuous settings: with
  normal martingales}, volume 1982 of \emph{Lecture Notes in Mathematics}.
\newblock Springer-Verlag, Berlin, 2009.

\bibitem[Ritchken(1995)]{ritchken}
P.~Ritchken.
\newblock On pricing barrier options.
\newblock \emph{Journal of Derivatives}, 3\penalty0 (2):\penalty0 19--28, 1995.

\bibitem[Sharpe(1978)]{sharpe}
W.F. Sharpe.
\newblock \emph{Investments}.
\newblock Prentice Hall, Englewood Cliffs, N.J., 1978.

\bibitem[Soroka(2006)]{soroka}
S.N. Soroka.
\newblock Good news and bad news: Asymmetric responses to economic information.
\newblock \emph{The Journal of Politics}, 68\penalty0 (2):\penalty0 372--385,
  2006.

\bibitem[Thompson(1977)]{wathompson}
W.A. Thompson.
\newblock On the treatment of grouped observations in life studies.
\newblock \emph{Biometrics}, 33\penalty0 (3):\penalty0 463--470, 1977.

\bibitem[Tian(1993)]{tian1993}
Y.~Tian.
\newblock A modified lattice approach to option pricing.
\newblock \emph{J. Futures Markets}, 13\penalty0 (5):\penalty0 563--577, 1993.

\bibitem[Tian(1999)]{tian1999}
Y.~Tian.
\newblock A flexible binomial option pricing model.
\newblock \emph{J. Futures Markets}, 19\penalty0 (7):\penalty0 817--843, 1999.

\bibitem[Uspensky(1937)]{uspensky}
J.V. Uspensky.
\newblock \emph{Introduction to mathematical probability}.
\newblock McGraw Hill, New York, 1937.

\bibitem[Walsh(2003)]{walsh}
J.B. Walsh.
\newblock The rate of convergence of the binomial tree scheme.
\newblock \emph{Finance Stoch.}, 7\penalty0 (3):\penalty0 337--361, 2003.

\bibitem[Williams(1991)]{williams}
D.~Williams.
\newblock \emph{Probability with martingales}.
\newblock Cambridge Mathematical Textbooks. Cambridge University Press,
  Cambridge, 1991.

\bibitem[Xiao(2010)]{xiao}
X.~Xiao.
\newblock Improving speed of convergence for the prices of european options in
  binomial trees with even numbers of steps.
\newblock \emph{Appl. Math. Comput.}, 216\penalty0 (9):\penalty0 2659--2670,
  2010.

\end{thebibliography}
\end{document}